\documentclass[aps,twocolumn,floatfix,nofootinbib]{revtex4}
\usepackage{amsmath,amsfonts,amssymb,graphicx,color,times,bbm}
\usepackage{epstopdf}
\usepackage{hyperref}
\usepackage{ulem}

\begin{document}

\bibliographystyle{apsrev}
\newtheorem{theorem}{Theorem}
\newtheorem{corollary}{Corollary}
\newtheorem{definition}{Definition}
\newtheorem{proposition}{Proposition}
\newtheorem{lemma}{Lemma}
\newcommand{\proofend}{\hfill\fbox\\\medskip }
\newcommand{\proof}[1]{{\bf Proof.} #1 $\proofend$}
\newcommand{\nn}{{\mathbbm{N}}}
\newcommand{\rr}{{\mathbbm{R}}}
\newcommand{\cc}{{\mathbbm{C}}}
\newcommand{\zz}{{\mathbbm{Z}}}
\newcommand{\mbp}{\ensuremath{\spadesuit}}
\newcommand{\je}{\ensuremath{\heartsuit}}
\newcommand{\jd}{\ensuremath{\clubsuit}}
\newcommand{\id}{{\mathbbm{1}}}
\renewcommand{\vec}[1]{\boldsymbol{#1}}
\newcommand{\me}{\mathrm{e}}
\newcommand{\mi}{\mathrm{i}}
\newcommand{\md}{\mathrm{d}}
\newcommand{\sg}{\text{sgn}}

\def\>{\rangle}
\def\<{\langle}
\def\({\left(}
\def\){\right)}

\newcommand{\ket}[1]{|#1\>}
\newcommand{\bra}[1]{\<#1|}
\newcommand{\braket}[2]{\<#1|#2\>}
\newcommand{\ketbra}[2]{|#1\>\!\<#2|}
\newcommand{\proj}[1]{|#1\>\!\<#1|}
\newcommand{\avg}[1]{\< #1 \>}

\renewcommand{\tensor}{\otimes}
\delimitershortfall=-2pt

\title{Spatial entanglement of bosons in optical lattices}

\author{M.\ Cramer,$^{1,2,}$\footnote{Correspondence and requests for materials should be addressed to M.C. (email: Marcus.Cramer@uni-ulm.de)} A.\ Bernard,$^{3}$ N.\ Fabbri,$^{3}$ L.\ Fallani,$^{3,4}$ C.\ Fort,$^{3}$ S.\ Rosi,$^{3}$ F.\ Caruso,$^{3,4}$ M.\ Inguscio,$^{3,4}$ and M.B.\ Plenio$^{1,2}$}

\affiliation{$^1$Institut f\"ur Theoretische Physik, Albert-Einstein Allee 11, Universit\"at Ulm, D-89069 Ulm, Germany \\
$^2$Center for Integrated Quantum Science and Technology, Albert-Einstein Allee 11, Universit\"at Ulm, D-89069 Ulm, Germany\\
$^3$LENS, Dipartimento di Fisica e Astronomia, Universit\`a di Firenze and INO-CNR, via Nello Carrara 1, I-50019 Sesto Fiorentino (FI), Italy, and\\
$^4$QSTAR, Largo Enrico Fermi 2, I-50125 Firenze, Italy}

\begin{abstract}
Entanglement is a fundamental resource for quantum information processing,
occurring naturally in many-body systems at low temperatures. The presence of entanglement
and, in particular, its scaling with the size of system partitions
underlies the complexity of quantum many-body states. The quantitative estimation of entanglement in
many-body systems represents a major challenge as it requires either full state
tomography, scaling exponentially in the system size, or the assumption
of unverified system characteristics such as its Hamiltonian or temperature. Here we adopt
recently developed approaches for the determination of rigorous lower entanglement bounds from readily accessible measurements
and apply them in an experiment of ultracold interacting bosons in optical lattices of approximately
$10^5$ sites. We then study the behaviour of spatial entanglement between the sites when crossing the superfluid-Mott
insulator transition and when varying temperature. This constitutes the first rigorous experimental large-scale entanglement quantification in a scalable quantum simulator.
\end{abstract}
\maketitle

\date{\today}

Entanglement plays a crucial role in most of the recent developments of quantum information processing and communication \cite{NC2000,PlenioV2007,HorodeckiHHH2009,GuhneToth09}.
Indeed, apart from the intrinsic interest in obtaining a deeper understanding of several counter-intuitive and surprising consequences of the quantum description of nature, it represents also a fundamental resource for various applications in quantum information science and metrology. From the practical point of view, once we manage to create such resource states, it is crucial to quantify the actual amount of entanglement contained in the created state to assess its degree of usefulness for quantum information processing protocols. This already challenging task becomes even more daunting in situations in which entanglement is shared between many different parties with the aim of implementing, for instance, multiparty quantum communication networks and distributed quantum computation.

\begin{figure}[t!]
\begin{center}
\includegraphics[width=1.\columnwidth]{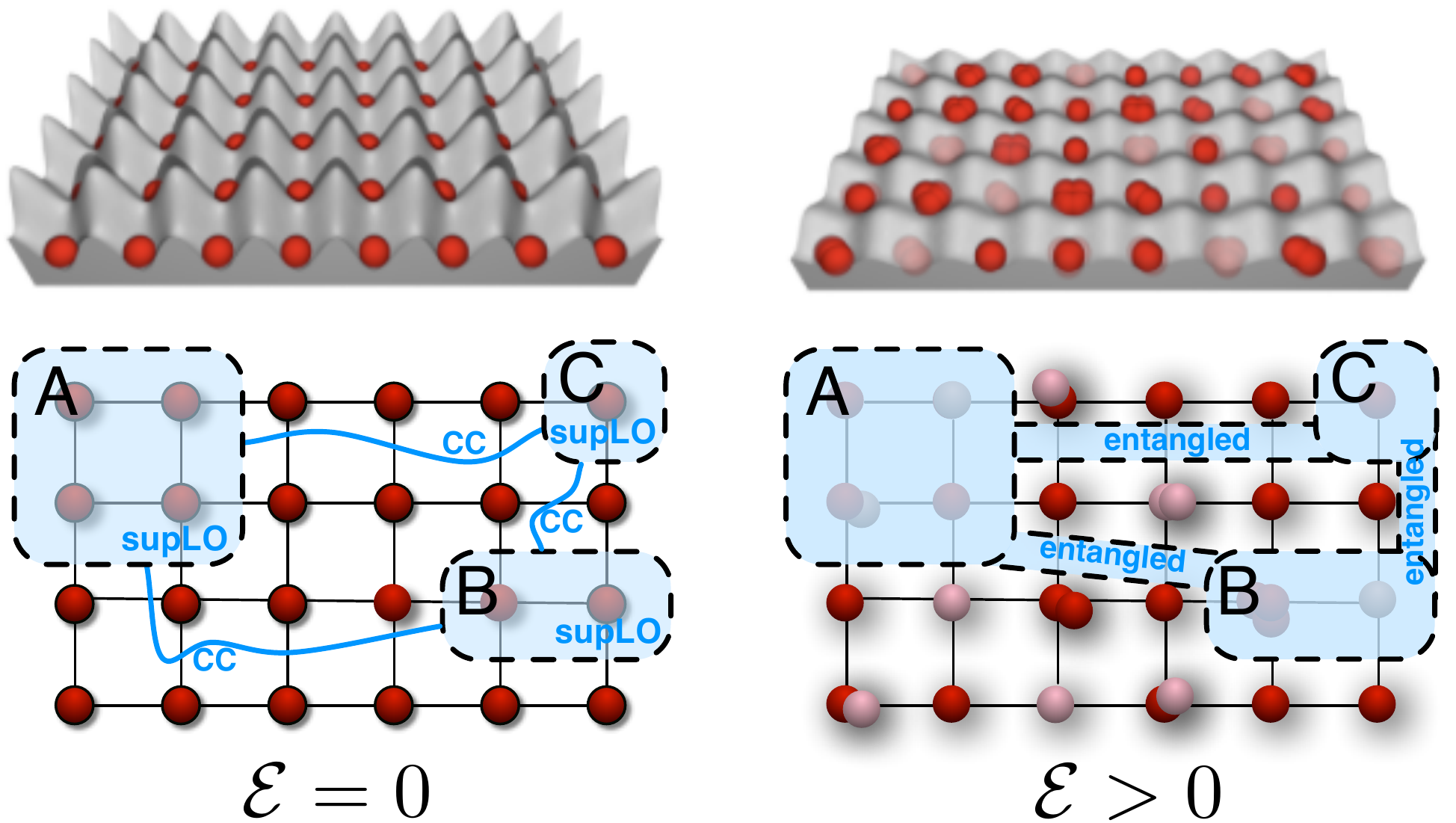}
\end{center}
\caption{{\bf Schematic illustration of the presence of entanglement}. We investigate entanglement in an optical lattice filled by ultracold bosons. At large lattice height, the ground state is a pure product state and no quantum correlations exist between any subset of sites (A, B, C, ...). Such a state can be created by only performing physical local operations (those that respect the mass superselection rule locally, supLO operations) and allowing the parties associated with the subsets to communicate classically (CC). Decreasing the lattice depth, the resulting state may not be created by such operations anymore -- the state becomes a resource of value $\mathcal{E}$ with which the parties may overcome their locality restrictions, i.e., entanglement is created.\label{cartoon}}
\end{figure}

In this respect, the last few years have seen experiments towards the verification of the presence of entanglement in a variety of physical realizations of many-body systems. Multi-particle spin entanglement of distinguishable particles has been created and studied experimentally for up to $14$ sites in ion traps \cite{Wineland6qubits,Haffner8qubits,Monz14qubits} and up to eight sites in photonic setups \cite{Yao2012} by means of entanglement witnesses that determine the presence  of entanglement. In ultracold neutral atomic gases, entanglement between indistinguishable particles with two internal degrees of freedom was generated by squeezing of the total (pseudo) spin and its presence verified by spin-squeezing inequalities \cite{EsteveGWGONature2008,LerouxSVPRL2010,RiedelBLHSTNature2010,Chauvet2010}.
For ultra-cold bosons with two internal degrees of freedom in optical lattices, entanglement between lattice sites was created by controlled collisions
and qualitative evidence for its presence was found \cite{MandelNature2003}. Experimental evidence for the entanglement between the spins in magnetic materials was given by comparison of the neutron scattering structure factor to a classical description \cite{Christensen2007}.

A key challenge that remains to be addressed however concerns the {\it quantitative} determination of the amount of entanglement. Even if the full state is known, the computation of its entanglement is a daunting task -- analytically and even numerically. If measurements are informationally incomplete, this is even impossible. Hence,  one needs to rely on the
determination of upper and lower bounds on the entanglement that is present in the system {\it without} resorting to assumptions concerning unverified system characteristics such as its Hamiltonian or its temperature.

We achieve this by adopting a simple but powerful principle \cite{AudenaertP2006,EisertBA2007, CramerPW2011,HorodeckiHH99,vollbrecht2007} in combination with methods from optimization theory: Given a set of observables,
we consider all density matrices that are compatible with experimentally measured expectation values of these observables. Amongst these density matrices, we find the one with the {\it least} amount of entanglement as
quantified by a suitable entanglement measure \cite{PlenioV2007,HorodeckiHHH2009}.
In this way we determine a lower bound on the entanglement that must have been present in the state that gave rise to the observed expectation values. These bounds do {\it not} require
any other assumptions on the system. We observe and quantify multi-partite entanglement in a periodic optical potential (optical lattice) that hosts
 bosonic atoms (see Fig.~\ref{cartoon}) and study its behaviour when crossing a quantum phase transition and when varying the temperature.

\begin{figure*}[t!]
	\begin{center}
		\includegraphics[width=1.\textwidth]{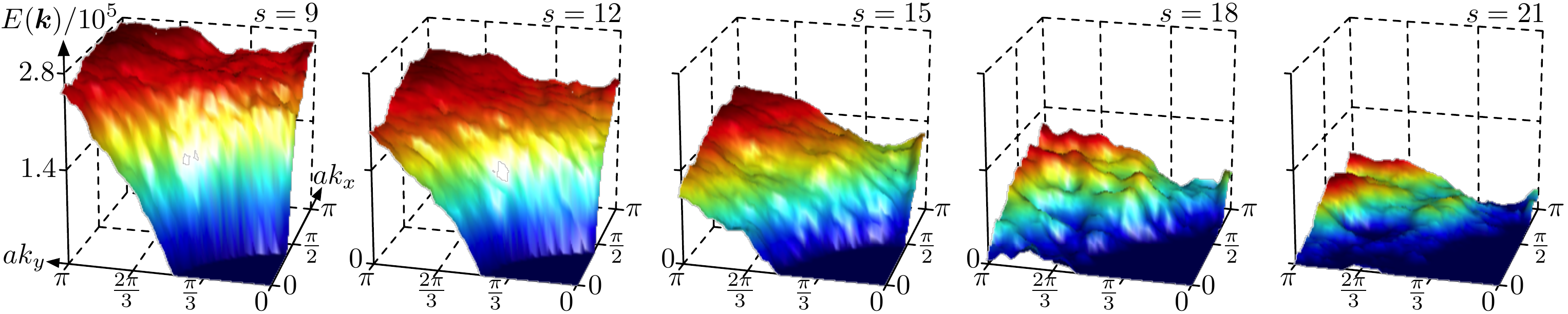}
	\end{center}
	\caption{{\bf Entanglement across a quantum phase transition}. The minimal entanglement $E(\vec{k})$ consistent with the measured momentum distributions of an atomic gas in a three-dimensional optical lattice	tuned across the quantum phase transition from superfluid to Mott insulator (lattice depth $s=9,12,15,18,21$). For each value of the momentum $\vec{k}$ in the first Brillouin zone, $E(\vec{k})$ provides a lower bound to the entanglement present in the bosonic sample. The total number of particles was approximately $3.5\times 10^5$ for each measurement. See Fig.~3 for total particle numbers.}
\label{Fig1}
\end{figure*}

\section{Results}

\subsection{Entanglement Quantification}

The entanglement that we quantify in our experiment is the entanglement of massive bosonic particles at different lattice sites. Hence, the non-local correlations are unavoidably intertwined with the superselection rules that prohibit the formation of coherent superpositions with different particle numbers. As a consequence, before we can provide a {\it rigorous} theoretical and experimental {\it quantification} of the entanglement in the system, we need to clarify its nature. We achieve this from the viewpoint of entanglement as a resource \cite{BrandaoP08}.

Physical constraints, fundamental or practical, impose limitations on accessible physical operations \cite{BrandaoP08,GourS08}. For example, the locality constraint that expresses the inability to exchange quantum particles between distant laboratories $A$ and $B$ prevents the execution of quantum gates between $A$ and $B$. Such constraints in turn imply the existence of resources---here entangled states---that, when consumed, allow for the realization of operations that are impossible under the given physical constraints \cite{Masanes2006}. The use of entanglement in teleportation for example allows for the realization of general quantum gates between $A$ and $B$ \cite{BennettBCJPW1993,EisertJPP2000}. This aspect captures the multi-partite nature and non-local correlations in the system.

For massive bosonic particles an even more fundamental constraint concerns superselection
rules for massive indistinguishable particles: Physical operations cannot create
coherent superpositions of different particle numbers. In connection with the locality constraint
this requires that all physically allowed local operations must commute with the local particle number operator.

The fact that both constraints, locality and super-selection rules, need to be considered {\it simultaneously} for massive bosonic particles leads to a refined picture of non-local resources and thus entanglement \cite{wiseman2003,verstraete2003,schuch2004,benatti2012}. In a multi-partite system, as in Fig.\ \ref{cartoon}, in which two or more
parties aim to exchange quantum information, but are restricted (i) to only act locally on their respective quantum system and communicate classically (LOCC), and (ii) to perform only operations preserving the local particle number operator, they may (in the two-partite case) only prepare states $\hat{\varrho}$ of the form
\begin{equation}
    \label{separable}
    \hat{\varrho}=\sum_np_n\hat{\varrho}^{(n)}_A\otimes\hat{\varrho}^{(n)}_B,
\end{equation}
where $\hat{\varrho}^{(n)}_A$ ($\hat{\varrho}^{(n)}_B$) are density operators of subsystem A (B), $\{p_n\}$ is a probability distribution and the local states must commute with the local particle number operators $\hat{N}_A$ and $\hat{N}_B$, respectively, i.e., $[\hat{\varrho}^{(n)}_A,\hat{N}_A] = [\hat{\varrho}^{(n)}_B,\hat{N}_B]=0$. All such states are here collected in the set $\mathcal{S}$. All other states, so states not in $\mathcal{S}$, become a resource, to be used to overcome the constraint imposed by locality and/or  superselection rules.

Before defining our entanglement quantifier, we illustrate the resource character of states that are not in $\mathcal{S}$, by means of an example taken from Ref.~\cite{verstraete2003}. Suppose a
single classical bit is encoded in the relative phase of the two states
\begin{equation}
\label{hidden}
    |\pm\rangle=\frac{1}{\sqrt{2}}\bigl(|0\rangle_A|1\rangle_B\pm |1\rangle_A|0\rangle_B\bigr).
\end{equation}
Remarkably, if Alice and Bob are constrained by LOCC {\it and} local particle number conservation, they are unable to distinguish these two states -- the bit $\pm$ is hidden from them. They may learn the bit, however, in one of two different ways: (i) if they share entanglement in the form
\begin{equation}
    \label{data_hiding_1}
    |\psi\rangle=\frac{1}{\sqrt{N+1}}\sum_{n=0}^N|n\rangle_A|N-n\rangle_B,
\end{equation}
which has a total number of particles $\langle\psi|(\hat{N}_A+\hat{N}_B)|\psi\rangle=\langle\psi|\hat{N}|\psi\rangle=N$, and $\text{tr}_B[|\psi\rangle\langle\psi|]$ and $\text{tr}_A[|\psi\rangle\langle\psi|]$ commute with the respective local number operators, but it may {\it not} be written as in Eq.~(\ref{separable}). {\it Or} (ii) if they share
the state ($z=|\alpha|\me^{\mi\phi}$)
\begin{equation}
    \label{data_hiding_2}
    \begin{split}
    \hat{\varrho}\propto\int_0^{2\pi}\!\!\!\md\phi\,|z\rangle_A\langle z|\otimes |z\rangle_B\langle
    z|,\;\;\;|z\rangle\propto\sum_{n=0}^\infty\frac{z^n}{\sqrt{n!}}|n\rangle,
\end{split}
\end{equation}
which is of the form in Eq.~(\ref{separable}), commutes with the total number operator and has $\langle\hat{N}\rangle=\text{tr}[\hat{N}\hat{\varrho}]=2|\alpha|^2$
but does {\it not} satisfy $[|z\rangle_A \langle z|,\hat{N}_A]=0$
.
Now, the success probability of learning the phase of $|\pm\rangle$ increases with $\langle\hat{N}\rangle$ (approaching unity as $\langle\hat{N}\rangle\rightarrow\infty$), or, in other words, the value of these resource states increases as their mean total number of particles with $\langle\hat{N}\rangle$ increases.

Let us finally consider an example from the context under experimental consideration. Bosons in optical lattices are well described by the Bose-Hubbard model, which is exactly solvabel in two extreme cases: (i) on-site interactions $U$ dominate over tunneling $J$ and (ii) the opposite case in which $J\gg U$. The ground state of this model in case (i) is simply a Fock state and thus a product state without any entanglement between sites. In the second case (ii), the model is equivalent to coupled harmonic oscillators and may be diagonalized by a symplectic transformation. Labeling sites by $i=1,\dots,L$, the ground state for fixed particle number $N$ reads
\begin{equation}
 \label{data_hiding_3}
\frac{(\sum_i\hat{b}_i^\dagger)^N}{\sqrt{L^NN!}}|\text{vac}\rangle=\sum_{\substack{n_1,\dots,n_L=0\\ \sum_in_i=N}}^Nc_{n_1,\dots,n_L}|n_1\cdots n_L\rangle,
\end{equation}
which may also serve as a resource to uncover the hidden bit in Eq.~(\ref{hidden}): In the bi-partite setting of two sites, $L=2$, following the protocol of Ref.~\cite{verstraete2003}, it may be shown that the success probability $p$ is given by
\begin{equation}
p=
\frac{1}{4}\sum_{n=1}^N|c_{n,N-n}+c_{n-1,N-n+1}|^2,
\end{equation}
which, as for the examples above, increases with $N$, tending to unity. This gives an example for the operational significance of the entanglement generated in optical lattices.

While for the above examples we considered a bi-partite setting, they readily carry over to a multi-partite setting \cite{verstraete2003}, which we will consider in the following.

The key point is now that we turn the above {\it qualitative} appreciation of the value of these states into a mathematically and physically well defined {\it quantifier} that may then be determined experimentally. It is a crucial requirement that the value of this quantifier does not increase on average under LOCC operations that respect local superselection rules (supLOCC in short). Such a  quantifier is then denoted a supLOCC monotone.

We start by defining the set $\mathcal{S}$ for an arbitrary partition. Let $\mathcal{L}_1\cup\cdots\mathcal{L}_P=\mathcal{L}$ be a $P$-partite partition of the lattice sites $\mathcal{L}$. A separable (with respect to this partition) state is then of the form
\begin{equation}
\label{P_separable}
\hat{\varrho}=\sum_np_n\,\hat{\varrho}^{(n)}_{\mathcal{L}_1}\otimes\cdots\otimes\hat{\varrho}^{(n)}_{\mathcal{L}_P},
\end{equation}
where $\hat{\varrho}^{(n)}_{\mathcal{L}_p}$ is the density operator corresponding to party (subsystem) $p$ and $\{p_n\}$ a probability distribution. The set $\mathcal{S}$ collects all states that are as in Eq.\ (\ref{P_separable}) and in addition commute with the local particle number, $[\hat{\varrho}^{(n)}_{\mathcal{L}_p},\sum_{\vec{i}\in\mathcal{L}_p}\hat{n}_{\vec{i}}]=0$. We now formulate the entanglement monotone for such arbitrary partitions and later, for the experiment, restrict our attention to the partition in which every lattice site constitutes a party. To this end, we define $\mathcal{W}$ as the set of entanglement witnesses $\hat{W}$ \cite{GuhneToth09} satisfying (i) $\text{tr}[\hat{W}\hat{\varrho}]\ge 0$ for all $\hat{\varrho}\in\mathcal{S}$ and (ii) the operator inequality $\hat{W}+\hat{N}\ge 0$. Then
\begin{equation}
    \label{monotone}
    \mathcal{E}(\hat{\varrho})=\max\bigl\{0,-\inf_{\hat{W}\in\mathcal{W}}\text{tr}[\hat{W}\hat{\varrho}]\bigr\}
\end{equation}
is a supLOCC monotone for {\it any} state $\hat{\varrho}$ (see Methods for a proof and note the similarity to entanglement monotones expressed as optimization over witnesses for spin systems in Ref.\ \cite{Brandao2005}). Note that, for any state $\hat{\varrho}$, its entanglement  $\mathcal{E}(\hat{\varrho})$ is upper bounded by the mean total number of particles $\langle\hat{N}\rangle=\text{tr}[\hat{N}\hat{\varrho}]$, providing a figure of merit for the lower bounds that we will present below.

We now set
out to quantify the entanglement under superselection rules contained in states of bosons
in optical lattices when each lattice site constitutes a party, i.e., labelling the sites
of the three-dimensional lattice by $\vec{i}=(i_1,i_2,i_3)$, states in $\mathcal{S}$ are of the
form $\sum_np_n\bigotimes_{\vec{i}}\hat{\varrho}_{\vec{i}}^{(n)}$ with $[\hat{\varrho}_{\vec{i}}^{(n)},\hat{n}_{\vec{i}}]=0$, where $\hat{n}_{\vec{i}}$ is the number operator for the lattice site $\vec{i}$. This corresponds to $\mathcal{L}=\cup_{\vec{i}\in\mathcal{L}}\{\vec{i}\}$ in the above definition. In this way, we quantify the entanglement shared between {\it sites} of the lattice (as opposed to between particles).

Needless to say, Eq.~(\ref{monotone}) is exceedingly hard to compute analytically or even
numerically, especially in the many-particle system (where each constituent is in addition being described in an infinite-dimensional Hilbert space) that we consider here. However, rather than aiming for exact values we will follow Refs.~\cite{AudenaertP2006,EisertBA2007,CramerPW2011,HorodeckiHH99,vollbrecht2007} to derive lower bounds to Eq.~(\ref{monotone}) which, after the introduction of essential aspects of the experimental setup, we will demonstrate to be obtained from readily accessible measurements.

\subsection{Experiment}
In the experiment we quantify the entanglement of a system of ultracold interacting bosons in a three dimensional lattice potential. An almost  pure\footnote{The estimated temperature of the condensate is $< 50$nK corresponding to a condensate fraction bigger than 80\%.} Bose-Einstein condensate  of $\approx 3.5 \times 10^5$ atoms of $^{87}$Rb  is prepared evaporating a sample of atoms in the $|F=1, M_F=-1\rangle$ state in a hybrid trap composed of a focused red detuned laser beam (optical dipole trap ODT) propagating in the horizontal plane plus a quadrupole magnetic field. As first demonstrated in \cite{Lin2009}, when the focused laser beam is slightly offset vertically from the center of the quadrupole magnetic field, the atoms experience a harmonic potential with cylindrical symmetry. In our system the resulting frequencies are $\approx$$50$~Hz and $\approx$$8$~Hz in the radial and axial directions respectively.
An optical lattice (OL) potential with lattice constant $a=\lambda/2$, generated by three counter-propagating red-detuned beams (with wavelength $\lambda=830.3$~nm and waists $w\approx180$~$\mu$m) is slowly superimposed to the sample by performing an exponential ramp in $t_\text{ramp}=140$~ms. The final amplitude of the lattices $V_{\text{OL}}=sE_\text{R}$ (where $E_\text{R}=h^2/(2m \lambda^2)$ is the recoil energy associated to the absorption of a lattice photon  by an atom with mass $m$) has been calibrated with an accuracy of $\pm10\%$ through lattice amplitude modulation spectroscopy \cite{Endres2011} and can be varied from $s=0$ to $s=30$.
In this way, we realize a many-body state of bosons in a three dimensional cubic lattice (subject to the harmonic trapping potential), the entanglement of which we are interested in quantifying.
The system's Hamiltonian is well approximated by the Bose-Hubbard Hamiltonian \cite{JackschBCGZ1998} (note that this information/approximation will not enter into our quantification of the entanglement). At sufficiently low temperatures, when the lattice depth is $s\gtrsim15$, the ratio between the interaction energy $U$ of two atoms in the same lattice site and the tunnel energy $J$ between two adjacent lattice sites is large enough to obtain a Mott insulator as firstly demonstrated in \cite{Greiner2002}.
Due to the overall harmonic confinement an inhomogeneous Mott insulator is obtained with regions with different filling. At $s=27$, we estimate $\sim$70\% of the atoms (in the outer region of the sample) to be in a Mott shell with 1 atom per site and $\sim$30\% (in the central region) to be in a Mott shell with 2 atoms per site.

We now describe the experimental procedure---time-of-flight measurements---which will give access to an observable allowing us to lower bound the entanglement of the bosons in the optical lattice. After a holding time $t_\text{hold}=5$~ms, we simultaneously switch off both the trap and the optical lattice. The cloud then expands freely for a time $t_\text{tof}=21$~ms  before we measure the column density through absorption imaging on a CCD camera. In practice, we  measure the optical density $\mu (x,y)$ of the atomic sample integrated along the direction of the imaging laser beam, and we extract the real atomic (column) density
\begin{equation}
    n(x,y)=\alpha \ (\mu(x,y)-\mu_0)
\end{equation}
with $\alpha$ being a pre-factor related to the imaging calibration and the effective size of the CCD square pixel, as discussed in the Methods section, and $\mu_0$ is the background noise level of the image, that is mainly due to residual fluctuations in the laser intensity of the imaging beam.
In the experiment, each value of the entanglement monotone is extracted from a set of about 40 atomic density profiles. The uncertainty associated to the monotone is estimated by adding systematic and statistical error in quadrature. The statistical contribution is given by the shot-to-shot fluctuation of the number of atoms recorded by each pixel of the CCD camera. Systematic contributions come from (a) the calibration of the absorption imaging efficiency (b) the estimation of the Wannier functions of the optical lattice.

\subsection{Experimental Entanglement Quantification}

From the column density $n(x,y)$ we can extract a lower bound to the entanglement $\mathcal{E}$ as follows.
The column density at position $(x,y)=\vec{r}= \hbar t \vec{k}/m$ (where $\vec{k}=(k_x\,k_y)$ is the quasi-momentum in the lattice) after a time of flight $t$ is given by \cite{pedri2001,gerbier2008}
\begin{equation}
    \hat{n}(\vec{k})=f(\vec{k})\sum_{\substack{\vec{i},\vec{j}\\
    i_z=j_z}}\hat{b}_{\vec{i}}^\dagger\hat{b}_{\vec{j}}\me^{a\mi \vec{k}\cdot
    (\vec{i}-\vec{j})}\me^{\mi\frac{m a^2}{2\hbar t} (\vec{j}^2-\vec{i}^2)},
    \label{momentum}
\end{equation}
where $\hat{b}_{\vec{i}}^\dagger$  ($\hat{b}_{\vec{i}}$) is the creation (annihilation) operator of a particle at
the lattice site $\vec{i}$ and $f(\vec{k})>0$ may be obtained from a numerical band--structure calculation (See Methods for details).

\begin{figure}[t!]
	\begin{center}
		\includegraphics[width=0.96\columnwidth]{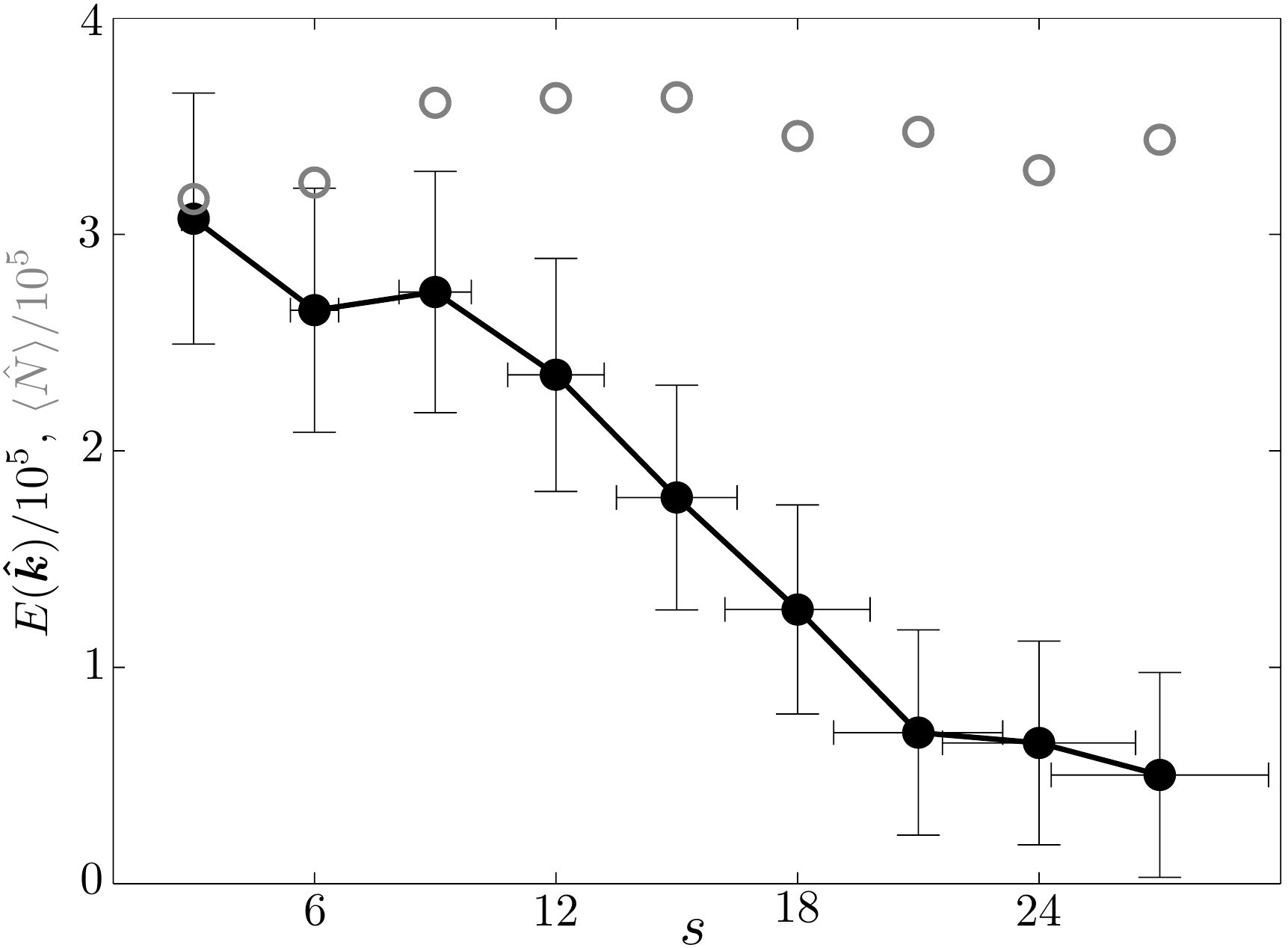}
	\end{center}
	\caption{{\bf Entanglement for different lattice depths}. Minimal entanglement $E(\vec{\hat{k}})$ consistent with time--of--flight measurements as a function  of the lattice depth $s$. The reported data correspond to the average of the lower bound $E(\vec{k})$ over $5\times 5$ pixels centered at $\vec{\hat{k}}$=($\pi/a$, $\pi/a$). Lines are guides to the eye. Error bars include statistical and
systematic errors due to  the uncertainty in the lattice depth and in the image acquisition and processing, see Methods for details. Circles in grey show the corresponding mean total number of bosons (without error bars for clarity, relative error $\approx 8\%$) which upper bounds the entanglement as discussed in the text after Eq.~(\ref{monotone}). }
\label{Fig2}
\end{figure}

It is straightforward to show \cite{CramerPW2011} that for the partition\footnote{In fact, it is also a lower bound for other partitions of the lattice $\mathcal{L}$: For any partition $\mathcal{L}=\cup_{p=1}^P\mathcal{L}_p$ with
$\text{tr}[\otimes_p\hat{\varrho}^{(n)}_{\mathcal{L}_p}\hat{b}_{\vec{i}}^\dagger\hat{b}_{\vec{j}}]=\text{tr}[\otimes_p\hat{\varrho}^{(n)}_{\mathcal{L}_p}\hat{b}_{\vec{i}}^\dagger]
\text{tr}[\otimes_p\hat{\varrho}^{(n)}_{\mathcal{L}_p}\hat{b}_{\vec{j}}]$
for $i_x\ne i_x$ or $i_y\ne j_y$, it constitutes a lower bound. Hence, e.g., it is also a lower bound for the entanglement between two-dimensional layers of the lattice, i.e., for the partition $\mathcal{L}=\cup_{i_x}\{(i_x,i_y,i_z)\}$. Another example being $\mathcal{L}=\cup_{i_x,i_y}\{(i_x,i_y,i_z)\}$, i.e., entanglement between chains arranged parallel to the  $z$-axis.}
$\mathcal{L}=\cup_{\vec{i}\in\mathcal{L}}\{\vec{i}\}$  the observable $\hat{W}(\vec{k}):=\hat{n}(\vec{k})/f(\vec{k})-\hat{N}\in\mathcal{W}$, where $\hat{N}$ is the total particle number operator, is a witness as required in Eq.~(\ref{monotone}) (see Refs.~\cite{CramerPW2011,KrammerKBBKM2009} for similar witnesses in the context of spin systems). Then, for any state $\hat{\varrho}$, we have a
lower bound to its entanglement content in terms of $\langle\hat{N}\rangle=\text{tr}[\hat{\varrho}\hat{N}]$ and $\langle\hat{n}(\vec{k})\rangle=\text{tr}[\hat{\varrho}\hat{n}(\vec{k})]$:
\begin{equation}
\label{lower_bound}
\mathcal{E}(\hat{\varrho})\ge \max\bigl\{0,\langle\hat{N}\rangle-\frac{\langle\hat{n}(\vec{k})\rangle}{f(\vec{k})}\bigr\}=:E(\vec{k}),
\end{equation}
which holds for all $\vec{k}$.
Note that there are no assumptions: The entanglement of {\it any} state is bounded from below by Eq.~(\ref{lower_bound}). The Hamiltonian governing the system, the temperature, details of external potentials, or even the system being in equilibrium, need not to be assumed.

As $E(\vec{k})$ is a lower bound to the entanglement $\mathcal{E}(\hat{\varrho})$ for all $\vec{k}$, averages
over an area $A$, $\int_{A}\md\vec{k}\,E(\vec{k})/|A|$, also provide lower bounds. We use this fact to account for the finite resolution of the camera and to
incorporate symmetries (see Methods for details). For ease of notation, we denote this lower bound also by $E(\vec{k})$.

To provide some examples of analytical calculations of $E(\vec{k})$ and to make the connection to the resource character in the data-hiding protocol above, let us evaluate the lower bound $E(\vec{k})$ for the class of states defined in the illustrative example above. In particular, suppose that the lattice consists of the two sites
$\vec{a}=(1\;0\;0)$
and $\vec{b}=(0\;1\;0)^t$, which we associate with Alice and Bob, respectively. For the states in Eqs.~(\ref{data_hiding_1}), (\ref{data_hiding_2}), and (\ref{data_hiding_3}), we find
\begin{equation}
E\bigl(\vec{k}=\tbinom{\pi/a}{0}\bigr)=
\begin{cases}
\frac{2}{N+1}\sum_{n=0}^{N-1}\sqrt{n+1}\sqrt{N-n},\\
2|\alpha|^2=\text{tr}[\hat{N}\hat{\varrho}],\\
N,
\end{cases}
\end{equation}
respectively. Hence, for all these states, we have a lower bound to the entanglement that is increasing in $\langle\hat{N}\rangle$ (thus capturing
the value of these states for uncovering the hidden bit in the above data-hiding protocol) and for the last two states the bound is in fact exact as
$\mathcal{E}(\hat{\varrho})\le \text{tr}[\hat{N}\hat{\varrho}]$ for all states $\hat{\varrho}$, showing the tightness of our bound.

Now we determine the minimal entanglement $E(\vec{k})$ for different optical lattice depths across the superfluid to Mott insulator transition. Before presenting our experimental results, let us consider the two extreme cases analytically. For ultra-deep lattices ($s\rightarrow\infty$) and at zero temperature, the system will be in a Fock state $|n_1n_2\dots\rangle$, for which $ \mathcal{E}=E(\vec{k})=0$ by definition. For very shallow lattices, when tunnelling dominates over the on-site repulsion and one neglects the latter, the ground state with $N$ particles of the translationally invariant Bose-Hubbard Hamiltonian is proportional to $(\sum_{\vec{i}}\hat{b}_{\vec{i}}^\dagger)^N|\text{vac}\rangle$ and one finds $ E(\vec{k})\approx N$ at $\vec{k}=\vec{\hat{k}}:=(\pi/a ,\pi/a)$. We thus expect the entanglement to decrease when increasing the lattice depth.

In Fig.~\ref{Fig1} we show $E(\vec{k})$ in the first Brillouin zone for $s=9,12,15,18,21$. For each value of $s$ we collected $\approx 40$ absorption images in order to reduce the statistical error on the determination of the entanglement. Relative shot--to--shot spread of the atom number is lower than $10\%$.
The lower bound of the entanglement  decreases as the system crosses the transition from the superfluid (lower $s$ values) to the Mott insulator phase ($s>15$). This behavior can be seen better in Fig.~\ref{Fig2}, reporting the value of $E(k_x,k_y)$ averaged over a box of $5 \times 5$ pixels around $\vec{\hat{k}}$, where we expect \cite{CramerPW2011} and found the bound to be largest. Details of our imaging system are explained in the Methods section, where we also describe the error analysis.

\begin{figure}[t!]
	\begin{center}
		\includegraphics[width=0.96\columnwidth]{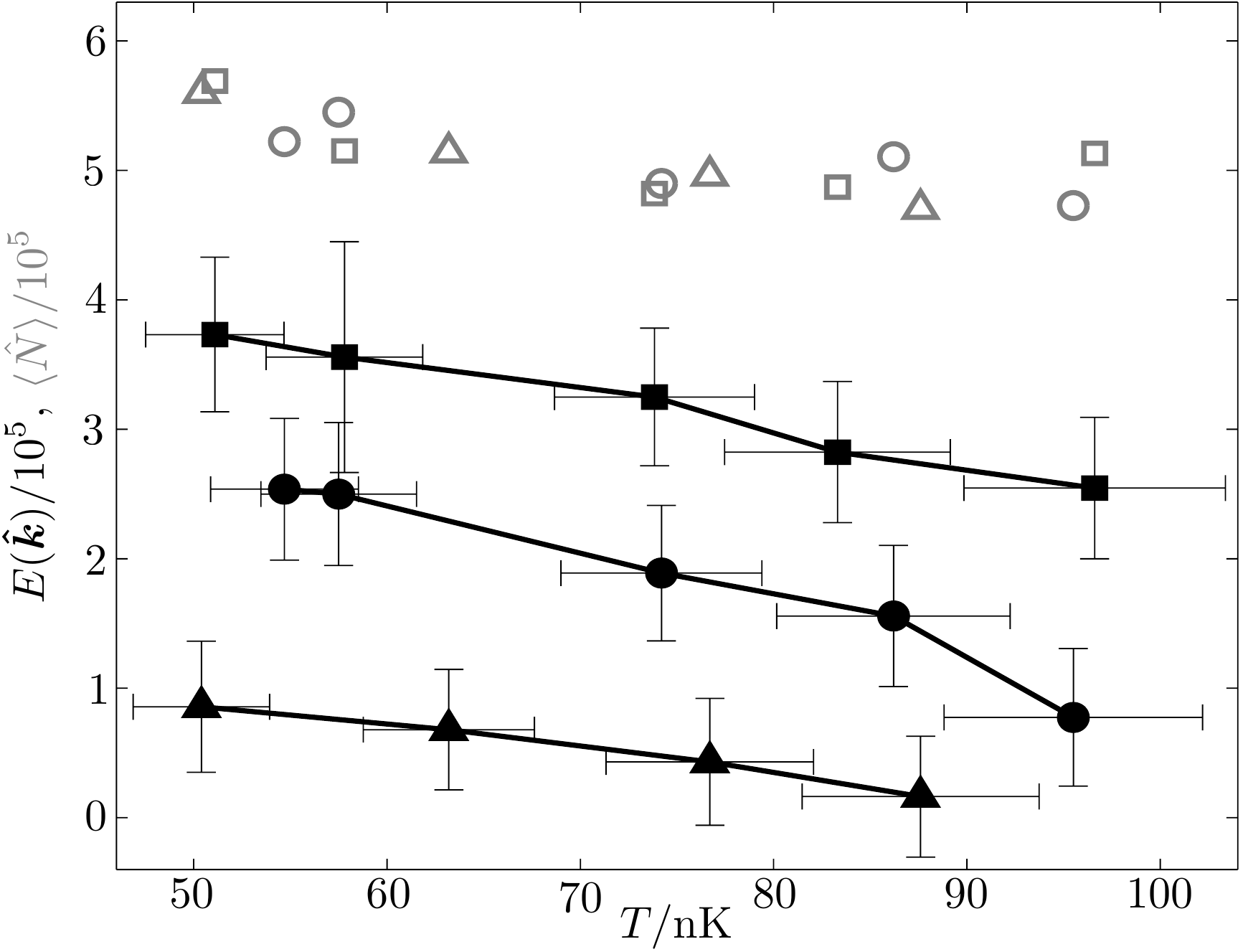}
	\end{center}
	\caption{{\bf Role of temperature in the entanglement behaviour}. Lower bound $E(\vec{\hat{k}})$ as a function of temperature for three different values of the lattice depth [$s=6$ (filled boxes), $s=12$ (filled circles), $s=18$ (filled triangles)] following the caption of Fig.~\ref{Fig2}. Corresponding non-filled symbols in grey show the mean total number of bosons.
Horizontal error bars indicate the uncertainty due to the calibration of the imaging system, vertical error bars as in Fig.~\ref{Fig2} (error bars on mean total number of atoms not shown for clarity, relative error was $\approx 8\%$).}
\label{Fig3}
\end{figure}

As the entanglement of the system is expected to decrease with increasing temperature \cite{CramerPW2011}, we also perform measurements fixing the optical lattice depth $s$ and varying the temperature of the atomic sample.
The determination of the temperature inside the lattice is still challenging \cite{Gemelke2009,Weld2009} while its measurement in a harmonic potential, i.e., before raising the lattice, is routinely done. Here, we refer to the temperature $T$ before the loading of the optical lattice. In practice, in order to realize samples of different temperatures, we perform optical evaporation in the ODT to different values of the power $P_{ODT}$ and then we increase the power of the optical dipole potential up to a fixed value $P_1$.
This procedure allowed us to obtain temperatures from $40$ nK up to $100$ nK in the same final harmonic potential before loading the lattices (see Methods for details).
In Fig.~\ref{Fig3} we show the behaviour of the minimal entanglement for different temperatures and for three different values of the lattice depth corresponding to a superfluid ground state ($s=6$), a Mott insulator phase ($s=18$), and the crossover region ($s=12$).  As expected \cite{CramerPW2011}, with increasing the temperature, the minimal entanglement consistent with the measurements decreases.

\section{Discussion}

We have quantified experimentally the multi-partite entanglement of a system of interacting bosons in an optical lattice through routinely done measurement of the atomic density profile after expansion. As the Hamiltonian and the ensuing dynamics of such a system can be controlled, it constitutes a bosonic {\it quantum simulator}, naturally supplying the resource entanglement at low temperatures. Our estimation of the entanglement is rigorous and without unspoken assumptions and provides a quantitative insight into the structure of the many--body state. In essence, we have answered the question, ``Which is the least amount of entanglement that is consistent with given measurements?" \cite{AudenaertP2006}.
The strategy we implemented for this estimation is sufficiently general to allow for its adaption and application in a wide variety of experimental settings that arise naturally in quantum science. Indeed, this principle may also be generalized to other quantities. One may
for example ask what is the maximal entropy consistent with given measurement results and by answering this question place rigorous, assumption-free, upper bounds on the entropy of a quantum many-body systems.
Of course, our approach is not restricted to bosonic systems but may also be applied to fermionic and spin systems or mixtures of bosonic and fermionic atoms, thus providing quantitative information about complex states of matter. It is the subject of ongoing research to directly relate notions of entanglement to the complexity of classical simulations. While in one-dimensional spin systems the connection between bi-partite entanglement and matrix product descriptions may be regarded as established \cite{mps_vs_area,area_review}, we hope that our work (in which we took the viewpoint of entanglement as a resource for quantum information tasks) inspires work towards this goal also in the massive-particle multi-partite setting.

\acknowledgements
The work at Ulm University has been supported by the EU Integrated Project QESSENCE,
the EU STREPs CORNER, the Alexander von Humboldt Professorship and the BMBF. The work at LENS has been supported by MIUR through PRIN nr. 2009TM7ERK\_004, ERC Advanced Grant DISQUA, EU FP7 Integrated Project AQUTE, and IIT Seed Project ENCORE.
The work of F.C. has been supported by EU FP7 Marie--Curie Programme
(Intra--European Fellowship and Career Integration Grant) and by MIUR--FIRB
grant (Project No. RBFR10M3SB). We thank Michele Modugno, Carlo Sias, Jianming Cai, and Gor Nikoghosyan for critical reading of the manuscript.
F.C. acknowledges H. Wunderlich for fruitful
discussions at the early stages of this project. The QSTAR is the MPQ, LENS, IIT, UniFi Joint Center for Quantum Science and Technology in Arcetri.

\begin{center}\bf\small Author contributions \end{center}
M.C., F.C. and M.B.P. proposed the project, M.C. and M.B.P. led the theory; A.B., N.F., L.F., C.F., S.R., and M.I. planned and carried out the experiment with input from F.C. and M.C.; M.C. derived the monotone and analyzed the data with input from all other authors; All authors discussed the results; M.C., M.B.P. (F.C., N.F., C.F.) wrote the theory (experimental) part of the manuscript with input from all authors.

$\,$

\section{Methods}

\subsection{A monotone under SSR-LOCC operations}
\label{appendix:monotone}
Here we show that
\begin{equation}
\mathcal{E}(\hat{\varrho})=\max\bigl\{0,-\inf_{\hat{W}\in\mathcal{W}}\text{tr}[\hat{W}\hat{\varrho}]\bigr\}
\end{equation}
is a monotone under LOCC operations commuting with the local particle number operators. Let the Hilbert space $\mathcal{H}$ be a direct product
of $\mathcal{G}$ parties, $\mathcal{H}=\bigotimes_{s=1}^{\mathcal{G}}\mathcal{H}_s$. LOCC operations with respect to this partition are operations taking density
matrices $\hat{\varrho}$ to $\sum_kp_k\hat{\varrho}_k$, where $\hat{\varrho}_k=\hat{A}_k\hat{\varrho}\hat{A}_k^\dagger/p_k$,
$p_k=\text{tr}[\hat{A}_k\hat{\varrho}\hat{A}_k^\dagger]$ and the $\hat{A}_k$ are of the form $\hat{A}_k=\bigotimes_{s=1}^{\mathcal{G}}\hat{A}_{s}^k$ and
fulfil $\sum_k\hat{A}_k^\dagger\hat{A}_k\le \id$ and $[\hat{A}_s^k,\hat{n}_s]=0$. $\mathcal{E}(\hat{\varrho})$ is an entanglement monotone if
\begin{equation}
\sum_kp_kE(\hat{\varrho}_k)
\le
\mathcal{E}(\hat{\varrho}).
\end{equation}
For all $\hat{W}\in\mathcal{W}$, we have $\text{tr}[\hat{W}\hat{\varrho}_k]\ge -\text{tr}[\hat{N}\hat{\varrho}_k]>-\infty$. Hence, the infimum
exists and we denote it by $E_k$. Now let $\epsilon>0$. Then $E_k+\epsilon$ is not an infimum and therefore there is an
$\hat{W}_{k,\epsilon}\in\mathcal{W}$ such that $\text{tr}[\hat{W}_{k,\epsilon}\hat{\varrho}_k]<E_k+\epsilon$, i.e.,
\begin{equation}
\label{xx}
\begin{split}
\sum_kp_kE(\hat{\varrho}_k)
&=-\sum_{\substack{k \\ E_k<0}}p_kE_k
<\sum_{\substack{k \\ E_k<0}}p_k(\epsilon-\text{tr}[\hat{W}_{k,\epsilon}\hat{\varrho}_k])\\
&\le \epsilon-\text{tr}\Bigl[\bigl(\sum_{k:\, E_k<0}\hat{A}_k^\dagger\hat{W}_{k,\epsilon}\hat{A}_k\bigr)\hat{\varrho}\bigr],
\end{split}
\end{equation}
which is upper bounded by
\begin{equation}
\sum_kp_kE(\hat{\varrho}_k)
\le \epsilon-\inf_{\hat{W}\in\mathcal{W}}\text{tr}[\hat{W}\hat{\varrho}],
\end{equation}
as the operator in brackets in Eq.\ (\ref{xx}) is a member of $\mathcal{W}$: For all $k$ and all $\epsilon$, we have
$\hat{W}_{k,\epsilon}+\hat{N}\ge 0$, i.e.,
\begin{equation}
\begin{split}
0&\le
\sum_{\substack{k \\ E_k<0}}\hat{A}_k^\dagger(\hat{W}_{k,\epsilon}+\hat{N})\hat{A}_k
\le \sum_{\substack{k \\ E_k<0}}\hat{A}_k^\dagger\hat{W}_{k,\epsilon}\hat{A}_k
+\hat{N}
\end{split}
\end{equation}
as $[\hat{A}_k,\hat{N}]=0$ for all $k$. Now let $\hat{\sigma}\in\mathcal{S}$. Then
\begin{equation}
\text{tr}\Bigl[\bigl(\sum_{\substack{k \\ E_k<0}}\hat{A}_k^\dagger\hat{W}_{k,\epsilon}\hat{A}_k\bigr)\hat{\sigma}\Bigr]
=\sum_{\substack{k \\ E_k<0}}\text{tr}[\hat{W}_{k,\epsilon}\hat{A}_k\hat{\sigma}\hat{A}_k^\dagger],
\end{equation}
where, up to normalization, $\hat{A}_k\hat{\sigma}\hat{A}_k^\dagger\in\mathcal{S}$, i.e., the above is lower bounded by zero (as
$\hat{W}_{k,\epsilon}\in \mathcal{W}$) and we hence have that for all $\epsilon>0$
\begin{equation}
\begin{split}
\sum_kp_kE(\hat{\varrho}_k)
< \epsilon-\inf_{\hat{W}\in\mathcal{W}}\text{tr}[\hat{W}\hat{\varrho}]
\le  \epsilon + E(\hat{\varrho}),
\end{split}
\end{equation}
which implies that $\mathcal{E}$ is an entanglement monotone.

\subsection{Density after time-of-flight}
\label{appendix:tof}
We set out to derive an expression of the atomic column density after free evolution, i.e.,
after evolution under the Hamiltonian
\begin{equation}
\label{free_ham}
\hat{H}=\int\!\md\vec{r}\,\hat{\Psi}^\dagger(\vec{r})\bigl[-\frac{\hbar^2}{2m}\vec{\nabla}^2\bigr]\hat{\Psi}(\vec{r}).
\end{equation}
 In order to connect the atomic column density to observables in the lattice, i.e., before the free expansion, we expand the field operators in Wannier functions
 of the lattice
 \begin{equation}
 \hat{\Psi}(\vec{r})=
 \sum_{\vec{i}}w_{\vec{i}}(\vec{r})\hat{b}_{\vec{i}}.
 \end{equation}
 Here $\vec{i}$ is a multi-index containing the lattice site and the band index.
 The density operator after evolution under $\hat{H}$ for a time $t$ reads
 \begin{equation}
 \begin{split}
 \hat{n}(\vec{r},t)& := \me^{\mi t\hat{H}/\hbar}\hat{\Psi}^\dagger(\vec{r})  \hat{\Psi}(\vec{r})\me^{-\mi t\hat{H}/\hbar}\\
 &=\sum_{\vec{i},\vec{j}}w^*_{\vec{i}}(\vec{r})w_{\vec{j}}(\vec{r})
 \me^{\mi t\hat{H}/\hbar}
 \hat{b}^\dagger_{\vec{i}}\hat{b}_{\vec{j}}\me^{-\mi t\hat{H}/\hbar}.
 \end{split}
 \end{equation}
 Now, due to the lattice geometry, the Wannier functions factorize and are the same for each spatial direction, $w_{\vec{i}}(\vec{r})=w_{i_x}(x)w_{i_y}(y)w_{i_z}(z)$. Owing to orthonormality, we hence find for the column-density operator $\hat{n}(x,y,t):=
 \int\md z\,\hat{n}(\vec{r},t)$ after time-of-flight~$t$
  \begin{equation}
  \nonumber
 \hat{n}(x,y,t)=\sum_{\substack{\vec{i},\vec{j}\\ i_z=j_z}}
 w^*_{i_x}(x)w_{j_x}(x)
 w^*_{i_y}(y)w_{j_y}(y)
 \hat{b}^\dagger_{\vec{i}}(t)\hat{b}_{\vec{j}}(t).
 \end{equation}
To compute the time-evolution $ \hat{b}_{\vec{i}}(t)=\me^{\mi t\hat{H}/\hbar}\hat{b}_{\vec{i}}\me^{-\mi t\hat{H}/\hbar}$ of the bosonic annihilation operators, we expand in orthonormal and complete plane waves $\omega(\vec{r})=\me^{2\pi\mi\vec{p}\cdot\vec{r}/L}/L^{3/2}$,
 \begin{equation}
 \hat{\Psi}(\vec{r})=
 \sum_{\vec{p}}\omega_{\vec{p}}(\vec{r})\hat{a}_{\vec{p}}.
 \end{equation}
 Due to orthonormality, the bosonic annihilation operators are related as
 \begin{equation}
 \label{relation}
 \begin{split}
 \hat{b}_{\vec{i}}= \sum_{\vec{p}}\langle w_{\vec{i}},\omega_{\vec{p}}\rangle\hat{a}_{\vec{p}},\;\;\;
  \hat{a}_{\vec{p}}=\sum_{\vec{i}}\langle \omega_{\vec{p}},w_{\vec{i}}\rangle \hat{b}_{\vec{i}},
 \end{split}
 \end{equation}
 where we denoted $\langle f,g\rangle=\int\md\vec{r}\,f^*(\vec{r})g(\vec{r})$.
 The Hamiltonian is diagonal in the basis of the $\hat{a}_{\vec{p}}$,
 \begin{equation}
 \hat{H}=\frac{2\pi^2\hbar^2}{L^2m}\sum_{\vec{p}}\vec{p}^2\hat{a}^\dagger_{\vec{p}}\hat{a}_{\vec{p}},
 \end{equation}
 which implies
 \begin{equation}
 \me^{\mi t\hat{H}/\hbar}\hat{a}_{\vec{p}}\me^{-\mi t\hat{H}/\hbar}=\me^{-\mi t\frac{2\pi^2\hbar}{L^2m}\vec{p}^2}\hat{a}_{\vec{p}},
 \end{equation}
 i.e., using Eqs.\ (\ref{relation}),
  \begin{equation}
  \nonumber
 \begin{split}
 \hat{b}_{\vec{i}}(t)
 &= \sum_{\vec{p}} \sum_{\vec{j}}\langle w_{\vec{i}},\omega_{\vec{p}}\rangle\langle \omega_{\vec{p}},w_{\vec{j}}\rangle\me^{-\mi t\frac{2\pi^2\hbar}{L^2m}\vec{p}^2}
 \hat{b}_{\vec{j}}.
 \end{split}
 \end{equation}
  Hence,
 \begin{widetext}
  \begin{equation}
 \begin{split}
 \hat{n}(x,y,t)
 &=
 \sum_{\substack{\vec{p},\vec{q},\vec{i},\vec{j}\\  i_z=j_z}}w^*_{i_x}(x)w_{j_x}(x)
 w^*_{i_y}(y)w_{j_y}(y)
 \langle \omega_{\vec{p}},w_{\vec{i}}\rangle\langle w_{\vec{j}},\omega_{\vec{q}}\rangle
  \me^{\mi t\frac{2\pi^2\hbar}{L^2m}(\vec{p}^2-\vec{q}^2)}
 \hat{a}^\dagger_{\vec{p}}\hat{a}_{\vec{q}},
 \end{split}
 \end{equation}
 where completeness and orthonormality imply $\sum_{i_z}\langle\omega_{p_z},w_{i_z}\rangle\langle w_{i_z},\omega_{q_z}\rangle=\delta_{p_z,q_z}$ and $\sum_{j}w_{j}(x)\langle w_{j},\omega_{q}\rangle=\omega_q(x)$, i.e.,
  \begin{equation}
  \hat{n}(x,y,t) =
 \sum_{\substack{\vec{p},\vec{q}\\  p_z=q_z}}
   \omega^*_{p_x}(x)  \omega_{q_x}(x)
 \omega^*_{p_y}(y)  \omega_{q_y}(y)
  \me^{\mi t\frac{2\pi^2\hbar}{L^2m}(\vec{p}^2-\vec{q}^2)}
 \hat{a}^\dagger_{\vec{p}}\hat{a}_{\vec{q}}.
  \end{equation}
  \end{widetext}

 Now, again using Eqs.\ (\ref{relation}), we arrive at an expression for the column density in terms of the $\hat{b}_{\vec{i}}$
\begin{equation}
  \label{tof_exact}
 \begin{split}
 \hat{n}(x,y,t)&=
 \sum_{\substack{\vec{i},\vec{j}\\ i_z=j_z}}
 g^*_{\vec{i}}(\vec{r},t)
 g_{\vec{j}}(\vec{r},t)\hat{b}^\dagger_{\vec{i}}\hat{b}_{\vec{j}},
 \end{split}
 \end{equation}
 where $g_{\vec{i}}(\vec{r},t)= g_{i_x}(x,t) g_{i_y}(y,t)$, with
 \begin{equation}
 \begin{split}
 g_{i}(x,t)&=\sum_{q} \omega_{q}(x)\langle \omega_{q},w_{i}\rangle \me^{-\mi t\frac{2\pi^2\hbar}{L^2m}q^2}\\
  &=\frac{1}{\sqrt{L}}\sum_{q} \me^{2\pi\mi qx/L}
  \langle \omega_{q},w_{i}\rangle \me^{-\mi t\frac{2\pi^2\hbar}{L^2m}q^2},
 \end{split}
 \end{equation}
 and we used completeness and orthonormality to arrive at $\sum_{p_z}\langle w_{i_z},\omega_{p_z}\rangle
\langle \omega_{p_z},w_{j_z}\rangle=\delta_{i_z,j_z}$.
 Finally, considering the properties of the Wannier functions, and after some algebra, we let $L\rightarrow\infty$ to obtain
 \begin{equation}
 \nonumber
 \begin{split}
 \frac{g_{i}(x,t)}{\sqrt{2\pi}}
 &=\me^{\mi \frac{\pi^2}{\tau}(x/a-i)^2}\int_{-\infty}^\infty\!\!\!\!\!\md\phi\, \me^{-\mi \tau\phi^2}\bar{w}_n(\phi+\tfrac{\pi}{a\tau}x-\tfrac{\pi}{\tau}i)\\
 &=:\me^{\mi \frac{\pi^2}{\tau}(x/a-i)^2}\frac{f_{i}(x)}{\sqrt{2\pi}},
 \end{split}
\end{equation}
 where  $\tau=t\frac{2\pi^2\hbar}{m a^2}$ and
\begin{equation}
  \bar{w}_n(\phi)=\frac{1}{\sqrt{2\pi}}
  \int_{-\infty}^{\infty}\md r\,w_{0,n}(ar)\me^{-2\pi\mi \phi r}
\end{equation}
is the Fourier transform of the Wannier function of the $n$'th band centred at zero. Eq.~(\ref{tof_exact}) with $g_{\vec{i}}$ as above is the exact expression for the column density at $(x\, y)$. All the involved functions may be obtained by a numerical calculation of the Wannier functions.

 In the experiment, only the lowest band is occupied and we omit the band index from now on.
 In the stationary phase approximation, which is valid for $1\ll\tau$ ($\approx 1.8\times 10^3$ in our experiment), one has
  \begin{equation}
 f_{i}(x)\approx (1-\mi)\frac{\pi}{\sqrt{\tau}}\bar{w}(\tfrac{ \pi}{a\tau}x-\tfrac{\pi}{\tau}i).
 \end{equation}
 Finally, approximating $\bar{w}(\tfrac{ \pi}{a\tau}x-\tfrac{\pi}{\tau}i)\approx \bar{w}(\tfrac{ \pi}{a\tau}x)$ \cite{gerbier2008}, yields
   \begin{equation}
   \nonumber
 \begin{split}
 \hat{n}(\vec{r}=\tfrac{\hbar t}{m}\vec{k},t)&=
 f(\vec{k})
 \sum_{\substack{\vec{i},\vec{j}\\ i_z=j_z\\ }}
\me^{\mi [ a\vec{k}(\vec{i}-\vec{j})
+ \pi^2( \vec{j}^2-\vec{i}^2)/\tau  ]}
 \hat{b}^\dagger_{\vec{i}}\hat{b}_{\vec{j}},
 \end{split}
 \end{equation}
 for the column density at $\vec{r}=(x\, y)=\hbar t (k_x\, k_y)/m=\hbar t \vec{k}/m=\tau a^2 \vec{k}/(2\pi^2)$ after a time-of-flight $t$. Here,
 \begin{equation}
  \label{observable2}
 \begin{split}
 f(\vec{k})&=\frac{m^2a^4}{\hbar^2 t^2}
 |w(\tfrac{ a}{2\pi}k_x)|^2|w(\tfrac{ a}{2\pi}k_y)|^2,\\
 w(\tfrac{ a}{2\pi}k)&=\frac{1}{\sqrt{2\pi}}
  \int_{-\infty}^{\infty}\md r\,w_{0}(ar)\me^{-\mi  k ar},
 \end{split}
 \end{equation}
 and $w_0$ is the Wannier function of the lowest band centred at zero.

\subsection{Image and error analysis}
\label{appendix:errors}
We intend now to analyze the measurement of
\begin{equation}
E(x,y)=
\langle\hat{N}\rangle-\frac{\langle\hat{n}(x,y)\rangle}{f(x,y)},
\end{equation}
where we now work in real-space coordinates, i.e., $\langle\hat{n}(x,y)\rangle=\int\md z\,\langle\hat{n}(x,y,z)\rangle$, where
$\langle\hat{n}(x,y,z)\rangle$ is the expectation value of the density distribution of the atom cloud at $\vec{r}=(x\, y\, z)$ after time-of-flight and
\begin{equation}
f(x,y)=\frac{m^2a^4}{\hbar^2 t^2}
 |w(\tfrac{ am}{2\pi\hbar t}x)|^2|w(\tfrac{ am}{2\pi\hbar t}y)|^2
\end{equation}
 as in Eq.~(\ref{observable2}).

Due to the spatial discretization of the CCD sensor used in the experiment, we define the discrete function
\begin{equation}
\begin{split}
E^\prime_{i,j}:&=\frac{1}{\Delta^2}
\int_{\Delta_i}\!\!\!\!\md x\int_{\Delta_j}\!\!\!\!\md y\, E(x,y)\\
&=\langle\hat{N}\rangle-\frac{1}{\Delta^2}
\int_{\Delta_i}\!\!\!\!\md x\int_{\Delta_j}\!\!\!\!\md y\,\frac{\langle\hat{n}(x,y)\rangle}{f(x,y)}
\end{split}
\label{Eij}
\end{equation}
where $(i,j)$ denotes the index of each pixel, centered on $(x_i, y_j)$, and  $\Delta_{i,j}=[x_{i,j}-\Delta/2,x_{i,j}+\Delta/2]$. $\Delta=2.78\,\mu$m is the effective pixel size which takes into account the physical pixel  size and the  magnification of the imaging system. The total number of atom is given by $N=\sum_{i,j} n_{i,j}$. Note that the quantity $E^\prime_{i,j}$ is still a lower bound for all $(i,j)$.

For each acquired image, we incorporate the symmetry of the observable
$\hat{n}(\vec{k})/f(\vec{k})$ by averaging over pixels corresponding to $(k_x,k_y)$, $(k_x\pm 2\pi/a,k_y\pm 2\pi/a)$, and the symmetry of the experimental setup by averaging also over the four points $(\pm k_x,\pm k_y)$. For Figs.~\ref{Fig2} and \ref{Fig3}, we additionally consider the average of $E^\prime_{i,j}$ on a subset of $5\times5$ pixels (corresponding to twice the width of the point spread function of the imaging system) centered around $\vec{k}\in [-2\pi/a,2\pi/a]^{\times 2}$, where $a$ is the lattice spacing. If we define a set of pixels $A$ over which we perform the average, the quantity
\begin{equation}
E_A^\prime:=\frac{1}{|A|}\sum_{(i,j)\in A}E_{i,j}^\prime
\,,
\end{equation}
is also a lower bound to the entanglement.

Actually, we do not have access to the quantity $n(x,y)/f(x,y)$ to be integrated in Eq.~(\ref{Eij}). Approximating (and taking the error into account below) the continuous function $f$ by $f(x,y)\approx f(x_i,y_j)$ for each $(x,y)\in(\Delta_i,\Delta_j)$, we introduce the simplified quantities $E_A$ and $E_{i,j}$
\begin{equation}
\begin{split}
E^\prime_{A}\approx E_{A}&:=\frac{1}{|A|}\sum_{(i,j)\in A} \left(\langle\hat{N}\rangle-\frac{1}{\Delta^2}
g_{i,j}\langle\hat{n}_{i,j}\rangle\right)\\
&=:\frac{1}{|A|}\sum_{(i,j)\in A}E_{i,j},
\end{split}
\label{Es}
\end{equation}
where $g_{i,j}=1/f(x_i,y_j)$ and $\langle\hat{n}_{i,j}\rangle$ is the expected number of atoms recorded by pixel $(i,j)$.

In the experiment, we use a running average of about 40 density profiles for each value of lattice depth $s$ and temperature $T$.
The empirical average of $E_A$ over a set of $M$ images is also the best estimation of the entanglement bound $E^\prime_A$
\begin{equation}
\begin{split}
\bar{E}_A&=\frac{1}{M}\sum_{n=1}^M E^{\,(n)}_A=\frac{1}{|A|}\sum_{(i,j)\in A}\bar{E}_{i,j},
\end{split}
\label{barEs}
\end{equation}
$n=1,\dots,M$ being the image index.
We proceed by analyzing the sources of uncertainty when estimating $E_A^\prime$, i.e., the systematic uncertainty related to $n_{i,j}$ and
$g_{i,j}$, and the statistical contribution associated to shot-to-shot variations of $n_{i,j}$.


The statistical uncertainty can be estimated as
\begin{equation}
\begin{split}
(\sigma^{stat}_{\bar{E}_A})^2&=\frac{1}{M(M-1)}\sum_{n=1}^M\left(E_A^{(n)}-\bar{E}_A\right)^2.
\end{split}
\end{equation}
The approximation $g(x,y)\simeq g_{i,j}$ used in
Eq.~(\ref{Es}) for  $x\in \Delta_i, y\in\Delta_j$ introduces a systematic error. We find
\begin{equation}
\bigl|E^\prime_{i,j}-E_{i,j}\bigr|
\le
\int_{\Delta_i}\!\!\!\!\md x\int_{\Delta_j}\!\!\!\!\md y\,
\frac{\langle\hat{n}(x,y)\rangle}{\Delta^2}\left|g(x,y)-g_{i,j}\right|.
\label{diff}
\end{equation}
From the mean value theorem we have that
\begin{equation}
\nonumber
g_{i,j}-g(x,y)=(\partial_x g)(a,b)(x_i-x)+(\partial_y g)(a,b)(y_i-y),
\end{equation}
where $(a,b)=(1-c)(x,y)+c(x_i,y_i)$ for some $c$ between $0$ and $1$. Hence, for $x\in\Delta_i$, $y\in\Delta_j$, we find
\begin{equation}
\nonumber
\begin{split}
|g_{i,j}-g(x,y)|&\le \frac{\Delta}{\sqrt{2}}\max_{a\in\Delta_i,b\in\Delta_j}
|\vec{\nabla}g(a,b)|
=:\frac{\Delta}{\sqrt{2}}\epsilon_{i,j},
\end{split}
\end{equation}
and with this result (\ref{diff}) becomes
\begin{equation}
\left|E^\prime_{i,j}-E_{i,j}\right|
\le
\frac{1}{\sqrt{2}\Delta}\epsilon_{i,j}\langle\hat{n}_{i,j}\rangle.
\end{equation}
Thus, assuming a flat error distribution, the resulting standard error is given by
\begin{equation}
\sigma_{i,j}=\frac{1}{\sqrt{6}\Delta}\epsilon_{i,j}
\langle\hat{n}_{i,j}\rangle.
\end{equation}


Now we analyze in more detail the other systematic contributions. In the following, we write $\vec{i}=(i\,j)$. The $g_{\vec{i}}$ are obtained from the Wannier function of the optical lattice and thus the error depends on the uncertainty $\sigma_s=0.1s$ we have in the estimation of the lattice depth $s$, i.e.,
\begin{equation}
\sigma_{g_{\vec{i}}}^2=\sigma_s^2(\partial_sg_{\vec{i}})^2.
\end{equation}

Now we discuss the systematic error on $n_{\vec{i}}^{(n)}$.
In the experiment, we measure the optical density distribution by absorption imaging. More specifically, we record on a CCD camera the intensity profile of a resonant probe laser beam interacting with the sample. The absorbed light intensity $I_a$, integrated  along the $z$ direction (the direction of propagation of the probe beam), as given by the Beer-Lambert law, is $I_a=I_0 \ \left(1-\exp[-\sigma \int{\hat{n}(\textbf{r}) dz}]\right)$, with $I_0$ being the incident intensity and $\sigma$ the resonant absorption cross section given by $\sigma=3\lambda_0^2/(2 \pi)$, where $\lambda_0$ is the wavelength of the resonant transition. For circularly-polarized light on the transition we used for $^{87}$Rb, we have $\sigma=2.907$~$10^{-13}$~m$^2$ \cite{Steck}. Hence, one has
\begin{equation}
n_{\vec{i}} = -\frac{1}{\sigma} \left(\frac{I_t - I_d}{I_0 - I_d} \right) \; ,
\end{equation}
where $I_t=I_0-I_a$ is the transmitted light intensity, and $I_d$ is the intensity of the background light recorded on the CCD camera without the imaging beam. Polarization effects and the atomic manifold level-structure of the optical transition used in the imaging process can bring the absorption cross-section to be smaller than its theoretical value given above. This would lead to underestimate the number of atoms. Thus, we performe an accurate calibration of the absorption imaging efficiency \cite{Reinaudi2007}.
For the pixel $i$ centered at $(x_i,y_i)$ as denoted above, the atomic density is given by
\begin{equation}
n_{\vec{i}}^{(n)} =\alpha(\mu_{\vec{i}}^{(n)}-\mu_0^{(n)}),
\end{equation}
where the prefactor $\alpha$ and its uncertainty $\sigma_{\alpha}$ are given by
\begin{equation}
\alpha=0.112,\;\;\;\sigma_\alpha=0.009,
\end{equation}
and $\mu_0^{(n)}$ is an offset that may vary from image to image (hence the index $n$).
We estimate $\mu_0^{(n)}$ as follows. For pixels $\vec{i}$ far away from the centre of the image, we do not expect any atoms. We consider
quadratic frames centred on the image of thickness one pixel and increasing size. We then calculate the average of $\mu_{\vec{i}}^{(n)}$ for
each frame and
take $\mu_0^{(n)}$ as the minimum over all such frames and ($F$ denotes the set of pixels corresponding to the frame)
\begin{equation}
\sigma^2_{\mu^{(n)}}=\frac{1}{|F|}\sum_{\vec{i}\in F}(\mu^{(n)}_{\vec{i}}-\mu_0^{(n)})^2.
\end{equation}


To summarize, the best estimation of the entanglement bound over a set of $M$ images is given by
\begin{equation}
\nonumber
\begin{split}
\bar{E}_A&=\frac{\alpha}{M}\sum_{n=1}^M\sum_{\vec{i}}(\mu_{\vec{i}}^{(n)}-\mu_0^{(n)})\left(1
-\frac{g_{\vec{i}}}{\Delta^2|A|}\delta_{\vec{i}\in A}\right),
\end{split}
\end{equation}
and our best estimation of the systematic uncertainty is
\begin{equation}
\begin{split}
(\sigma^{sys}_{\bar{E}_A})^2&=\sigma_\alpha^2(\partial_\alpha\bar{E}_A)^2+\sum_{n=1}^M\sigma_{\mu_0^{(n)}}^2(\partial_{\mu_0^{(n)}}\bar{E}_A)^2
\\
&\hspace{3cm}+\sum_{\vec{i}\in A}\sigma_{g_{\vec{i}}}^2(\partial_{g_{\vec{i}}}\bar{E}_A)^2,
\end{split}
\end{equation}
which evaluates to
\begin{equation}
\nonumber
\begin{split}
(\sigma^{sys}_{\bar{E}_A})^2&=\frac{\sigma_\alpha^2\bar{E}^2_A}{\alpha^2}+\frac{\alpha^2}{M^2}\left(
\sum_{\vec{i}\in A}\frac{g_{\vec{i}}}{\Delta^2|A|}
-\sum_{\vec{i}}1
\right)^2\sum_{n=1}^M\sigma_{\mu_0^{(n)}}^2
\\
&\hspace{3cm}+\frac{1}{\Delta^4|A|^2}\sum_{\vec{i}\in A}\sigma_{g_{\vec{i}}}^2\bar{n}^2_{\vec{i}}.
\end{split}
\end{equation}
The global uncertainty can be found adding systematic and statistical errors in quadrature,
\begin{equation}
\sigma_{E_A^\prime}^2=
(\sigma^{sys}_{\bar{E}_A})^2+(\sigma^{stat}_{\bar{E}_A})^2+\frac{1}{|A|^2}\sum_{(i,j)\in A}\sigma_{i,j}^2,
\label{err}
\end{equation}
which corresponds to the error bars indicated in the main text.
Analyzing the different contributions, we have found that the main error sources are related to the estimation of $\alpha$ and $\mu_0^{(n)}$, hence the error bars actually do not decrease when considering a larger number of images, i.e. statistical errors are negligible. In other words, our entanglement estimation does not require a large set of absorption images in order to get small errors on the lower bound.

\subsection{Production of samples with different temperatures}

\label{appendix:temperature}
\begin{figure}[t]
\begin{center}
\includegraphics[width=0.95\columnwidth]{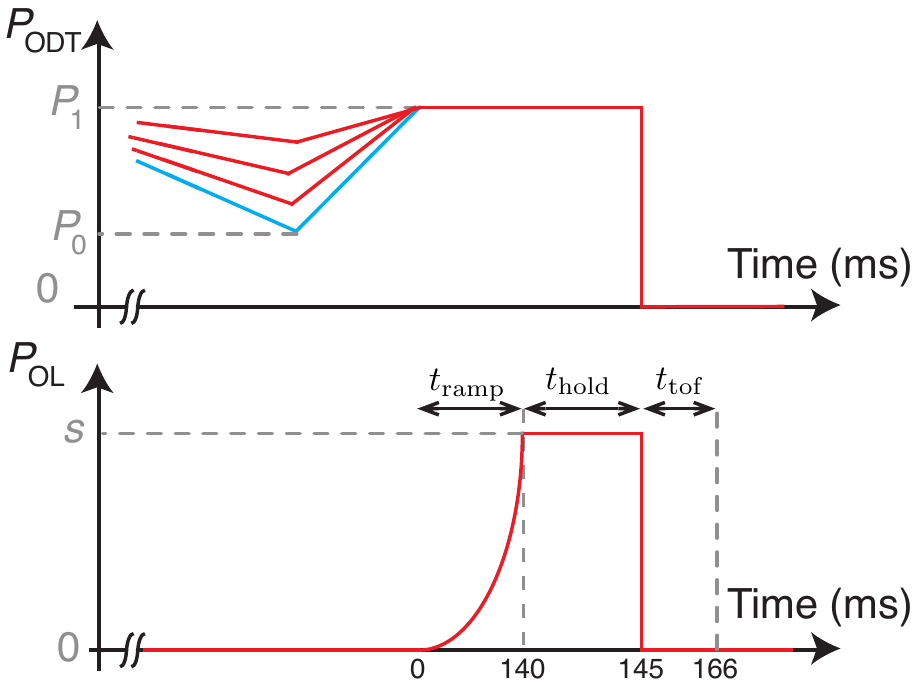}
\end{center}
\caption{{\bf Varying the temperature of the sample}. Experimental sequence to obtain samples with different temperatures in the same final potential before ramping the lattices: Ramps of the optical dipole trap (ODT) and the optical lattice (OL). The value of $P_0$ corresponds to the power of the ODT at the end of the evaporation. It is tuned in order to realize samples of different temperatures.}
\label{ramps}
\end{figure}

In Fig.~\ref{ramps} we show the experimental sequence adopted to obtain samples with different temperature $T$ (the temperature before ramping up the lattice) in the same harmonic potential. As the temperature is varied through a final evaporation performed lowering the power $P_{\text{ODT}}$ of the optical dipole trap
to a final value $P_0$, in general samples with different temperatures are obtained in different harmonic potentials. For this reason, before ramping up the lattices we  adiabatically increase in 500~ms the ODT to a fixed power $P_1$. In this way all the samples with different temperatures are prepared in the same harmonic potential with cylindrical symmetry characterized by a radial frequency of $50$ Hz and an axial frequency of $8$ Hz.
The atom number in the samples with different temperatures is kept constant by varying the loading time of the magnetic trap in order to realize samples with similar atom number.

\end{document}